\documentclass[times, 10pt]{article} 
\begin{document}

\begin{center}

\begin{Large}
%%% Scientific Data Mining in Astronomy
SCIENTIFIC DATA MINING IN ASTRONOMY
\end{Large}

\bigskip

\begin{large}
%%% \chapterauthor{Kirk D. Borne}{George Mason University}
Kirk D. Borne
\end{large}

\bigskip

{\it{Department of Computational and Data Sciences,
George Mason University,
Fairfax, VA 22030, USA\\
kborne@gmu.edu}}

\end{center}

\bigskip
\bigskip

%%% Scientific Data Mining in Astronomy
%%% \bigskip

\parskip0pt

\begin{center}
{\large{Abstract}}
\end{center}

\begin{quote}
\begin{small}
We describe the application of 
data mining algorithms to research problems
in astronomy.  We posit  that data mining
has always been fundamental to astronomical
research, since data mining 
is the basis of evidence-based discovery,
including classification, clustering,
and novelty discovery.
These algorithms represent a major
set of computational tools for discovery in large
databases, which will be increasingly essential
in the era of data-intensive astronomy.  Historical
examples of data mining in astronomy are reviewed, 
followed by a discussion of one of the largest 
data-producing projects anticipated for the coming decade:
the Large Synoptic Survey Telescope (LSST).  To facilitate
data-driven discoveries in astronomy, we envision
a new data-oriented research paradigm for 
astronomy and astrophysics -- astroinformatics.
Astroinformatics is described as both a research
approach and an educational imperative for modern
data-intensive astronomy.  An important application area
for large time-domain sky surveys (such as LSST)
is the rapid identification, characterization,
and classification of real-time sky events
(including moving objects, photometrically
variable objects, and the appearance of 
transients).  We describe one possible implementation
of a classification broker for such events, which
incorporates several astroinformatics techniques:
user annotation, semantic tagging, metadata markup, 
heterogeneous data integration, 
and distributed data mining.
Examples of these types of collaborative classification
and discovery approaches
within other science disciplines are presented.
\end{small}
\end{quote}

\section{Introduction}
\label{s:intro}

It has been said that astronomers have been doing
data mining for centuries: {\it{``the data are mine,
and you cannot have them!''}}.  Seriously, astronomers
are trained as data miners, because we are trained to: 
(a)~characterize the known 
({\it{i.e.}}, unsupervised learning, clustering);
(b)~assign the new
({\it{i.e.}}, supervised learning, classification);
and 
(c)~discover the unknown
({\it{i.e.}}, semi-supervised learning, outlier detection)
\cite{borne01A, borne01B}.
These skills are more critical than ever since 
astronomy is now a data-intensive science, and it will
become even more data-intensive in the coming decade
\cite{brunner2001, szalay2002, becla06}.

We describe the new data-intensive research paradigm 
that astronomy and astrophysics are now entering
\cite{djorgovski01c, borne2009a, borne2009b}. 
This is described within the context of the largest 
data-producing astronomy project in the coming decade –- 
the LSST (Large Synoptic Survey Telescope). 
The enormous data output, database contents, 
knowledge discovery, and community science expected 
from this project will impose massive data challenges 
on the astronomical research community.  One of these 
challenge areas is the rapid machine learning (ML), 
data mining, and classification of all novel 
astronomical events from each 3-gigapixel (6-GB) 
image obtained every 20 seconds throughout every night 
for the project duration of 10 years.  We describe these 
challenges and a particular implementation of a 
classification broker for this data fire hose.
But, first, we review some of the prior results of
applying data mining techniques in astronomical research.
A similar, more thorough survey of data mining and ML 
in astronomy was published \cite{ballbrunner2009}
after this paper was published\footnote{
Borne, K.,
``Scientific Data Mining in Astronomy,''
in Next Generation of Data Mining (Taylor \& Francis: CRC Press), 
pp. 91-114 (2009)}.

\section{Data Mining Applications in Astronomy}
\label{s:dmr}

Astronomers classically have focused on clustering and
classification problems as standard practice in our
research discipline.   This is especially true
of observational (experimental) astronomers who
collect data on objects in the sky, and then try to
understand the objects' physical properties and hence understand
the underlying physics that leads to those properties.
This invariably leads to a partitioning of the objects
into classes and subclasses, which reflect the 
manifestation of different physical processes
that appear dominant in different classes of objects.
Even theoretical astrophysicists, who apply pure physics
and applied mathematics to astronomy problems, are 
usually (though not always) governed by the results
of the experimentalists -- to identify classes of
behavior within their models, and to make predictions
about further properties of those classes that will
enhance our understanding of the underlying physics.

\subsection{Clustering}
Clustering usually has a 
very specific meaning to an astronomer -- that is ``spatial
clustering'' (more specifically, angular clustering 
on the sky).  In other words, we see groupings of stars
close together in the sky, which we call star clusters. 
We also see groupings of galaxies in the sky, which we call
galaxy clusters (or clusters of galaxies).  On even
larger spatial scales, we see clusters of clusters of
galaxies (superclusters) -- e.g.,
our Milky Way galaxy belongs to the Local Group of Galaxies,
which belongs to the Local Supercluster.  
Most of these cluster classes can be further subdivided
and specialized: e.g., globular star clusters 
versus open star clusters; or loose groups of galaxies
versus compact groups of galaxies; or rich clusters of
galaxies versus poor clusters of galaxies.  Two of the
research problems that are addressed by astronomers
who study these objects are discovery and membership -- i.e., 
discovering new clusters, and assigning objects as
members of one or another cluster.   These astronomical
applications of clustering are similar to
corresponding ML applications.
Because clustering is standard research practice
in astronomy, it is not possible
to summarize the published work in this area, since it
would comprise a significant fraction of all research
papers published in all astronomy journals and
conference proceedings over the last century.  
Specific data mining applications of clustering in
astronomy include the search for rare and new types
of objects \cite{djorgovski01a, djorgovski01b, dutta07}.

More generally, particularly for the ML 
community, clustering refers to class discovery and segregation
within any parameter space (not just spatial clustering). 
Astronomers perform this general type of clustering also
\cite{fundplane, djorgovski01c}.   
For example, there are many objects
in the Universe for which at least two classes have
been discovered.  Astronomers have not been too
creative in labeling these classes, which include: 
Types I and II supernovae,
Types I and II Cepheid variable stars, Populations
I and II (and maybe III) stars, Types I and II 
active galaxies, and so on, including further refinement 
into subclasses for some of these.  These observationally
different types of objects (segregated classes) were 
discovered when astronomers noticed clustering
in various parameter spaces (i.e., in scatter plots
of measured scientific parameters).

\subsection{Classification}

The other major dimension of astronomical research
is the assignment of objects to classes.  This was
historically carried out one-at-a-time, as the data
were collected one object at a time.  ML
and data mining classification algorithms were not
explicitly necessary.  However, in fact, the process
is the same in astronomy as in data mining: 
(1)~class discovery (clustering);
(2)~discover rules for the different classes
(e.g., regions of parameter space); 
(3)~build training samples to refine the rules; 
(4)~assign new objects to known classes using
new measured science data for those objects.
Hence, it is accurate to say that astronomers have
been data mining for centuries.  Classification is
a primary feature of astronomical research.  We are
essentially zoologists -- we classify objects in
the astronomical zoo.

As the data sets have grown in size, it has become
increasingly appropriate and even imperative to
apply ML algorithms to the data in order
to learn the rules and to apply the rules of classification.
Algorithms that have been used include Bayesian analysis,
decision trees, neural networks, 
and (more recently) support vector machines.
The ClassX project has used a network of classifiers 
in order to estimate classes of X-ray sources using
distributed astronomical data collections \cite{class-x, class-x2}.

We will now briefly summarize some specific examples of these.
But first we present a more general survey of data mining
research in astronomy.

\subsection{General Survey of Astronomical Data Mining}
\label{s:astrodm}

A search of the online astronomical literature database
ADS (NASA's Astrophysics Data System) lists only 63 refereed
astronomy research papers (765 abstracts of all types --
refereed and unrefereed)
that have the words ``data mining'' 
or ``machine learning'' in their abstracts.  
(Note that ADS searches a much broader set of disciplines
than just astronomy when non-refereed papers are included --
most of these search results are harvested from the ArXiv.org 
manuscript repository.)
Of course, there are many fine papers related
to astronomical data mining
in the SIAM, ACM, IEEE, and other journals
and proceedings that are not harvested by ADS.

Within the ADS list of {\it{refereed}} papers, 
the earliest examples that explicitly refer 
to ``data mining'' in their abstract are two papers
that appeared in 1997 -- these were general perspective
papers.  (Note that there are many papers,
including \cite{borne01A} and \cite{borne01B}
that were not in the refereed literature
but that pre-date the 1997 papers.)
The first of the refereed data mining application papers
that explicitly mentions ``data mining'' in the abstract
and that focused on a specific astronomy research problem
appeared in 2000 \cite{bohdan2000}.  This paper
described all-sky monitoring and techniques 
to detect millions of variable (transient)
astronomical phenomena of all types.  This was an excellent
precursor study to the LSST (see \S \ref{s:lsst-db}).

Among the most recent examples of refereed papers in ADS
that explicitly refer to data mining (not including this author's
work \cite{borne08}) is the paper \cite{mahabal2008} that addresses 
the same research problem as \cite{bohdan2000}: the automated
classification of large numbers of transient and variable objects.
Again, this research is a major contributor and precursor
to the LSST research agenda (\S \ref{s:lsst-dm}).

A very recent ``data mining'' paper focuses on automatic prediction 
of Solar CMEs (Coronal Mass Ejections), which lead to energetic
particle events around the Earth-Moon-Mars environment,
which are hazardous to astronauts outside the protective
shield of the Earth's magnetosphere \cite{qahwaji2008}.
This is similar to the data mining research project
just beginning at George Mason University 
with this author \cite{olmeda2008}.

Additional recent work includes investigations into
robust ML for terascale astronomical datasets
\cite{ball2007a}, 

In addition to these papers, several astronomy-specific
data mining projects are underway.  These include
AstroWeka (http://astroweka.sourceforge.net/),
Grist (Grid Data Mining for Astronomy; http://grist.caltech.edu),
the Laboratory for Cosmological Data Mining (http://lcdm.astro.uiuc.edu/), 
the LSST Data Mining Research study group (\cite{borne2008b}),
the Transient Classification Project at Berkeley \cite{bloom2008},
and the soon-to-be commissioned Palomar Transient Factory.

We will now look at more specific astronomical applications
that employed ML and data mining techniques.
We have not covered everything (e.g., other methods that
have been applied to astronomical data mining include
principal component analysis, kernel regression, random forests,
and various nearest-neighbor methods, such as 
\cite{whitmore1984, fundplane, ferreras, bornechang, carliles2007, ball2007b}).

\subsubsection{Bayesian Analysis}

A search of ADS lists 575 refereed papers
in which the words Bayes or Bayesian appear in the
paper's abstract.  For comparison, the same search
criteria returned 2313 abstracts of all papers (refereed
and non-refereed).  Seven of the refereed papers were
published before 1980 (none published before 1970).
One of these was by Sebok \cite{sebok79}.  
He applied Bayesian probability analysis to the most
basic astronomical classification problem -- 
distinguishing galaxies from stars among 
the many thousands of objects detected in large images.
This is a critical problem in astronomy, since the study
of stars is a vastly different astrophysics regime than
the study of galaxies.  To know which objects in the
image are stars, and hence which objects are galaxies,
is critical to the science.  It may seem that this is an obvious 
distinction, but that is only true for nearby galaxies,
which appear large on the sky (with large angular extent).
This is not true at all for very distant galaxies,
who provide the most critical information about the origin 
and history of our Universe.  These distant galaxies appear
as small blobs on images, almost indistinguishable 
from stars -- nearly 100\% of the stars are in our 
Milky Way Galaxy, hence very very nearby (by astronomical
standards), and consequently stars therefore
carry much less cosmological significance.

A more recent example is the application of Bayesian
analysis to the problem of star formation in young
galaxies \cite{kampak08}: the authors applied a Bayesian
Markov Chain Monte Carlo method to determine whether
the stars in the galaxies form in one
monolithic collapse of a giant gas cloud,
or if they form in a hierarchical fashion (with stars
forming in smaller galaxies forming first, then those galaxies 
merge to become larger galaxies, and so on).  
The latter seems to be the best model to fit the
observational data.

The above examples illustrate a very important point.
The large number of papers that refer to Bayes analysis
does not indicate the number that are doing data mining.
This is because Bayesian analysis is used primarily
as a statistical analysis technique or as a probability
density estimation technique.  The latter is certainly
applicable to classification problems, but not on a grand scale
as we expect for data mining (i.e., discovering hidden
knowledge contained in large databases).  

One significant recent paper that applies Bayesian analysis 
in a data mining sense focuses on a very important problem
in large-database astronomy: cross-identification of
astronomical sources in multiple large data collections
\cite{budavari2007}.  In order to match the same object
across multiple catalogs, the authors have proposed the use
of more than just spatial coincidence, but also include
numerous physical properties, including colors, redshift
(distance), and luminosity.  The result is an efficient
algorithm that is ready for petascale astronomical data mining.

\subsubsection{Decision Trees}

A search of ADS lists 21 refereed papers (166 
abstracts of all types) in which ``Decision Tree''
appears in the abstract.  One of the earliest 
(non-refereed) conference
papers was the 1994 paper by Djorgovski, Wier, and Fayyad
\cite{skicat1994}, when Fayyad was working at NASA's 
Jet Propulsion Lab.  This paper described the SKICAT
classification system, which was the standard example
of astronomical data mining quoted in many 
data mining conference talks subsequently.
The earliest paper we could find in astronomy was in 1975
\cite{howard1975},
16 years before the next paper appeared.  The 1975 paper
addressed a {\it{``new methodology to integrate planetary 
quarantine requirements into mission planning, 
with application to a Jupiter orbiter''.}}

Decision trees have been applied to another critical research
problem in astronomy by \cite{salzberg95} -- the identification
of cosmic ray (particle radiation) contamination in
astronomical images.  Charge-coupled device (CCD) 
cameras not only make excellent
light detectors, they also detect high-energy particles
that permeate space.  Cosmic-ray particles deposit their energy
and create spikes in CCD images (in the same way
that a light photon does).  The cosmic-ray hits
are random (as the particles enter the detector randomly
from ambient space) -- they have nothing to do with the
image.  Understanding the characteristics of these
bogus ``events'' (background noise)
in astronomical images and being able
to remove them are very important steps in astronomical
image processing.  The decision tree classifiers employed by
\cite{salzberg95} produced 95\% accuracy.  Recently, researchers
have started to investigate the application of neural
networks to the same problem \cite{waniak06} (and others).

\subsubsection{Neural Networks}

A search of ADS lists 418 refereed papers
in which the phrases ``Neural Net'' or ``Neural Network'' appear in the
paper's abstract.  For comparison, the same search
criteria returned over 10,000 abstracts of all papers (refereed
and non-refereed, most of which are not in astronomy; 
see \S \ref{s:astrodm}).  The earliest of these refereed papers
that appeared in an
astronomical journal \cite{jeffrey86}
(published in 1986) addressed neural networks and simulated
annealing algorithms in general.  One of the first real
astronomical examples that was presented in a refereed paper
\cite{angel90} applied a neural network to the problem
of rapid adaptive mirror adjustments in telescopes in order
to dramatically improve image quality.

As mentioned above, artificial neural networks 
(ANN) have been applied to the 
problem of cosmic-ray detection CCD images.  ANN
have also been applied to another important problem
mentioned earlier: star-galaxy discrimination (classification)
in large images.  Many authors have applied ANN
to this problem, including 
\cite{odewahn92, odewahn93, bazell98, mahonen2000, andreon2000, 
cortiglioni2001, philip2002, qin2003}.  
Of course, this astronomy research
problem has been tackled by many algorithms, including
decision trees \cite{ball2006}.

Two other problems that have received a lot of 
astronomical research attention using neural networks
are: (a)~the classification of different galaxy types
within large databases of galaxy data
(e.g., \cite{storrie92, naim95, goderya2002, ball2004});
and (b)~the determination of the photometric redshift
estimate, which is used as an approximator of 
distance for huge numbers of galaxies, for which
accurate distances are not known
(e.g., \cite{firth2003, collister2004, vanzella2004, oyaizu2008}).  
The latter problem has also been
investigated recently using random forests \cite{carliles2007}
and support vector machines.

\subsubsection{Support Vector Machines (SVM)}

ADS lists 154 abstracts (refereed and non-refereed) that
include the phrase ``Support Vector Machine'', of which 21 of these
are refereed astronomy journal papers.  Three of the latter
focus on the problem mentioned earlier: determination of 
the photometric redshift estimate for distant galaxies
\cite{wadadekar2005, way2006, wang2008}. 
Note that \cite{wang2008} also applies a kernel regression method
to the problem -- the authors find that kernel
regression is slightly more accurate than SVM, but
they discuss the positives and negatives of the two methods.
SVM was used in conjunction with a variety of other methods
to address the problem of cross-identification of
astronomical sources in multiple data collections that 
was described earlier \cite{rohde05, rohde06}.
SVM has also been used by several authors for 
forecasting solar flares and solar wind-induced geostorms, 
including \cite{gavri2001, qu2006, qahwaji2008}.

\section{Data-Intensive Science}
\label{s:dis}

The development of models to describe and understand 
scientific phenomena has historically proceeded at a 
pace driven by new data.  The more we know, the more 
we are driven to tweak or to revolutionize our models, 
thereby advancing our scientific understanding. 
This data-driven modeling and discovery linkage has 
entered a new paradigm \cite{mah07}.  The acquisition of 
scientific data in all disciplines is now accelerating 
and causing a nearly insurmountable data avalanche \cite{bell05}. 
In astronomy in particular, rapid advances in three 
technology areas (telescopes, detectors, and computation) 
have continued unabated –- all of these advances lead to 
more and more data \cite{becla06}. With this accelerated advance in 
data generation capabilities, humans will require novel, 
increasingly automated, and increasingly more effective 
scientific knowledge discovery systems \cite{borne06}. 

To meet the data-intensive research challenge, the 
astronomical research community has embarked on a 
grand information technology program, to describe 
and unify all astronomical data resources worldwide.  
This global interoperable virtual data system is referred 
to as the National Virtual Observatory 
(NVO, at www.us-vo.org) in the U.S., 
or more simply the ``Virtual Observatory'' (VO). Within the 
international research community, the VO effort is steered 
by the International Virtual Observatory Alliance 
(IVOA at www.ivoa.net). 
This grand vision encompasses more than a collection of 
data sets. The result is a significant evolution in the 
way that astrophysical research, both observational and 
theoretical, is conducted in the new millennium \cite{mcdow04}.  
This revolution is leading to an entirely new branch of 
astrophysics research –- {\it{Astroinformatics}} -– still in its 
infancy, consequently requiring further research and 
development as a discipline in order to aid in the 
data-intensive astronomical science that is emerging \cite{borneast06}.
	
The VO effort enables discovery, access, and integration 
of data, tools, and information resources across all 
observatories, archives, data centers, and individual 
projects worldwide \cite{plante04}.  However, it remains outside 
the scope of the VO projects to generate new knowledge, 
new models, and new scientific understanding from the 
huge data volumes flowing from the largest sky survey 
projects [8, 9].  Even further beyond the scope of the 
VO is the ensuing feedback and impact of the potentially 
exponential growth in new scientific knowledge discoveries 
back onto those telescope instrument operations.  
In addition, while the VO projects are productive 
science-enabling I.T. research and development projects, 
they are not specifically {\it{scientific}} 
(astronomical) research projects.  
There is still enormous room for scientific data portals 
and data-intensive science research tools that integrate, 
mine, and discover new knowledge from the vast distributed 
data repositories that are now VO-accessible \cite{borne06,
borne2009a}.

The problem therefore is this: astronomy researchers 
will soon (if not already) lose the ability to keep up 
with any of these things: the data flood, the scientific 
discoveries buried within, the development of new models 
of those phenomena, and the resulting new data-driven 
follow-up observing strategies that are imposed on 
telescope facilities to collect new data needed to 
validate and augment new discoveries.

\section{Astronomy Sky Surveys as Data Producers}
\label{s:surveys}

A common feature of modern astronomical sky surveys is 
that they are producing massive (terabyte) databases.  
New surveys may produce hundreds of terabytes (TB) up to 
100 (or more) petabytes (PB) both in the image data archive 
and in the object catalogs (databases). Interpreting these 
petabyte catalogs ({\it{i.e.}}, mining the databases for new 
scientific knowledge) will require more sophisticated 
algorithms and networks that discover, integrate, and 
learn from distributed petascale databases more effectively
\cite{gray2002}, \cite{longo2001}.

\subsection{The LSST Sky Survey Database}
\label{s:lsst-db}

One of the most impressive astronomical sky surveys 
being planned for the next decade is the Large Synoptic 
Survey Telescope project (LSST at www.lsst.org) \cite{tyson04}.  
The three fundamental distinguishing astronomical attributes 
of the LSST project are: 

\begin{enumerate}

\item {\it{Repeated temporal measurements}} of all observable 
objects in the sky, corresponding to thousands of observations 
per each object over a 10-year period, expected to generate 
10,000-100,000 alerts each night –- an alert is a signal 
({\it{e.g.}}, XML-formatted news feed) to the astronomical research 
community that something has changed at that location on the sky: 
either the brightness or position of an object, or the 
serendipitous appearance of some totally new object;

\item {\it{Wide-angle imaging}} that will repeatedly cover 
most of the night sky within 3 to 4 nights (= tens of 
billions of objects); and 

\item {\it{Deep co-added images}} of each observable 
patch of sky (summed over 10 years: 2016-2026), 
reaching far fainter objects and to greater distance 
over more area of sky than other sky surveys \cite{strauss04}. 

\end{enumerate}

Compared to other astronomical sky surveys, the LSST survey 
will deliver time domain coverage for orders of magnitude more 
objects. It is envisioned that this project will 
produce $\sim$30 TB of data per each night of observation for 10 years.  
The final image archive will be $\sim$70 PB (and
possibly much more), and the final LSST 
astronomical object catalog (object-attribute database) 
is expected to be $\sim$10-20 PB.  

LSST's most remarkable data product will be a 10-year ``movie'' 
of the entire sky = ``{\it{Cosmic Cinematography}}''.
This time-lapse coverage of the night sky will open up
time-domain astronomy like no other project has been able to do
previously.   In general, astronomers
have a good idea of what things in the sky
are varying and what things are not varying, as a result 
of many centuries of humans staring at the sky, with and without 
the aid of telescopes. But, there is so much more possibly happening 
that we are not aware of at the very faintest limits simply because 
we have not explored the sky systematically night after night on 
a large scale. When an unusual time-dependent event occurs in the sky 
({\it{e.g.}}, a gamma-ray burst, supernova, or in-coming asteroid), 
astronomers (and others) will not only want to examine spatial 
coincidences of this object within the various surveys, 
but they will also want to search for other data covering that 
same region of the sky that were obtained at the same time as this 
new temporal event. These contextual data will enable
more robust classification and characterization
of the temporal event. Because of the time-criticality and potential 
for huge scientific payoff of such follow-up observations of
transient phenomena, the classification system must also be able 
to perform time-based searches very efficiently and very effectively 
({\it{i.e.}}, to search all of the distributed VO databases 
as quickly as possible).
One does not necessarily know in advance if such a new discovery
will appear in any particular waveband, and so one will want to examine 
all possible astronomical sky surveys for coincidence events.
Most of these ``targets of opportunity'' will consequently be added 
immediately to the observing programs of many ground-based 
and space-based astronomical telescopes, observatories,
and on-going research experiments worldwide.

\subsection{The LSST Data-Intensive Science Challenge}

LSST is not alone.  It is one (likely the biggest one) 
of several large astronomical sky survey projects 
beginning operations now or within the coming decade.  
LSST is by far the largest undertaking, in terms of 
duration, camera size, depth of sky coverage, volume 
of data to be produced, and real-time requirements on 
operations, data processing, event-modeling, and 
follow-up research response.  One of the key features 
of these surveys is that the main telescope facility 
will be dedicated to the primary survey program, with 
no specific plans for follow-up observations.   This 
is emphatically true for the LSST project \cite{mould04}.  
Paradoxically, the follow-up observations are 
scientifically essential –- they contribute 
significantly to new scientific discovery, to the 
classification and characterization of new astronomical 
objects and sky events, and to rapid response to 
short-lived transient sky phenomena.

Since it is anticipated that LSST will generate many 
thousands (probably tens of thousands) of new astronomical 
event alerts per night of observation, there is a critical 
need for innovative follow-up procedures.  These procedures 
necessarily must include modeling of the events –- 
to determine their classification, time-criticality, 
astronomical relevance, rarity, and the scientifically 
most productive set of follow-up measurements.  Rapid 
time-critical follow-up observations, with a wide range 
of time scales from seconds to days, are essential for 
proper identification, classification, characterization, 
analysis, interpretation, and understanding of nearly 
every astrophysical phenomenon ({\it{e.g.}}, supernovae, novae, 
accreting black holes, microquasars, gamma-ray bursts, 
gravitational microlensing events, extrasolar planetary 
transits across distant stars, new comets, incoming 
asteroids, trans-Neptunian objects, dwarf planets, 
optical transients, variable stars of all classes, 
and anything that goes ``bump in the night'').

\subsection{Petascale Data Mining with the LSST}
\label{s:lsst-dm}

LSST and similar large sky surveys have enormous potential 
to enable countless astronomical discoveries.  
Such discoveries will span the full spectrum of statistics: 
from rare one-in-a-billion (or one-in-a-trillion) 
type objects, to a complete statistical and astrophysical 
specification of a class of objects (based upon millions 
of instances of the class).  One of the key scientific 
requirements of these projects therefore is to learn 
rapidly from what they see. This means: (a)~to identify 
the serendipitous as well as the known; (b)~to identify 
outliers ({\it{e.g.}}, ``front-page news'' discoveries) that 
fall outside the bounds of model expectations; 
(c)~to identify rare events that our models say should be there; 
(d)~to find new attributes of known classes; 
(e)~to provide statistically robust tests of existing models; 
and (f)~to generate the vital inputs for new models.  
All of this requires integrating and mining of all known data: 
to train classification models and to apply classification models.

LSST alone is likely to throw such data mining and knowledge 
discovery efforts into the petascale realm.  For example: 
astronomers currently discover $\sim$100 new supernovae 
(exploding stars) per year. Since the beginning of human 
history, perhaps $\sim$10,000 supernovae have been recorded.  
The identification, classification, and analysis of supernovae 
are among the key science requirements for the LSST Project 
to explore Dark Energy -– {\it{i.e.}}, supernovae contribute to 
the analysis and characterization of the ubiquitous cosmic 
Dark Energy.  Since supernovae are the result of a rapid 
catastrophic explosion of a massive star, it is imperative 
for astronomers to respond quickly to each new event with 
rapid follow-up observations in many measurement modes 
(light curves; spectroscopy; images of the host galaxy's 
environment).  Historically, with $<$10 new supernovae 
being discovered each week, such follow-up has been feasible.  
But now, LSST promises to produce a list of 
{\it{1000 new supernovae each night}} for 
10 years \cite{strauss04}, which represent a small fraction of the 
total (10-100 thousand) alerts expected each night! 
Astronomers are faced with the enormous challenge of 
efficiently mining, correctly classifying, and 
intelligently prioritizing a staggering number of 
new events for follow-up observation each night for a decade.

The major features and contents of the LSST scientific database include: \\
---  $>$100 database tables \\
---  Image metadata = 675M rows \\
---  Source catalog = 260B rows \\
---  Object catalog = 22B rows, with 200+ attributes \\
---  Moving Object catalog \\
---  Variable Object catalog \\
---  Alerts catalog \\
---  Calibration metadata \\
---  Configuration metadata \\
---  Processing metadata \\
---  Provenance metadata \\

Many possible scientific data mining use cases
are anticipated with the LSST database, including:

\begin{itemize}

\item Provide rapid probabilistic classifications for all 
10,000 LSST events each night;

\item Find new ``fundamental planes'' of parameters ({\it{e.g.}}, 
the fundamental plane of Elliptical galaxies);

\item Find new correlations, associations, relationships 
of all kinds from 100+ attributes in the science database;

\item Compute N-point correlation functions over a variety 
of spatial and astrophysical parameters;

\item Discover voids or zones of avoidance in multi-dimensional 
parameter spaces ({\it{e.g.}}, period gaps);

\item Discover new and exotic classes of astronomical objects,
while discovering new properties of known classes;

\item Discover new and improved rules for classifying known 
classes of objects ({\it{e.g.}}, photometric redshifts);

\item Identify novel, unexpected behavior in the time domain 
from time series data of all known variable objects;

\item Hypothesis testing -- verify existing (or generate new) 
astronomical hypotheses with strong statistical confidence, 
using millions of training samples;

\item Serendipity --  discover the rare one-in-a-billion type 
of objects through outlier detection; and

\item Quality assurance --  identify glitches, anomalies, 
image processing errors through deviation detection.

\end{itemize}

Some of the data mining research challenge areas posed
by the petascale LSST scientific database include:

\begin{itemize}

\item scalability (at petabytes scales) of existing ML 
and data mining algorithms;

\item development of grid-enabled parallel data mining algorithms;

\item designing a robust system for brokering classifications from 
the LSST event pipeline;

\item multi-resolution methods for exploration of petascale databases;

\item visual data mining algorithms for visual exploration of the massive databases;

\item indexing of multi-attribute multi-dimensional astronomical 
databases (beyond sky-coordinate spatial indexing); and

\item rapid querying of petabyte databases.

\end{itemize}

\section{A Classification Broker for Astronomy}
\label{s:broker}

We are beginning to assemble user requirements and design 
specifications for a ML engine (data integration 
network plus data mining algorithms) to address the petascale 
data mining needs of the LSST and other large data-intensive 
astronomy sky survey projects. The data requirements surpass 
those of the current Sloan Digital Sky Survey (SDSS,
at www.sdss.org) by 1000-10,000 
times, while the time-criticality requirement 
(for event/object classification and characterization) 
drastically drops from months (or weeks) down to minutes 
(or tens of seconds). In addition to the follow-up 
classification problem (described above), astronomers 
also want to find every possible new scientific 
discovery (pattern, correlation, relationship, outlier, 
new class, etc.) buried within these new enormous databases. 
This might lead to a petascale data mining compute engine 
that runs in parallel alongside the data archive, testing 
every possible model, association, and rule.  We will 
focus here on the time-critical data mining engine 
({\it{i.e.}}, classification broker) that enables rapid follow-up 
science for the most important and exciting astronomical 
discoveries of the coming decade, on a wide range of time 
scales from seconds to days, corresponding to a plethora 
of exotic astrophysical phenomena.

\subsection{Broker Specifications: AstroDAS}

The classification broker's primary specification is to 
produce and distribute scientifically robust near-real-time 
classification of astronomical sources, events, objects, or 
event host objects ({\it{i.e.}}, the astronomical object
that hosts the event; {\it{e.g.}}, the host galaxy for
some distant supernova explosion -- it is important to
measure the redshift distance of the host galaxy in order
to interpret and to classify properly the supernova). 
These classifications are derived from 
integrating and mining data, information, and knowledge 
from multiple distributed data repositories. The broker 
feeds off existing robotic telescope and astronomical 
alert networks world-wide, and then integrates existing 
astronomical knowledge (catalog data) from the VO. The 
broker may eventually provide the knowledge discovery 
and classification service for LSST, a torrential fire 
hose of data and astronomical events. 

Incoming event alert data will be subjected to a suite of 
ML algorithms for event classification, 
outlier detection, object characterization, and novelty discovery.  
Probabilistic ML models will produce rank-ordered lists of the 
most significant and/or most unusual events.  These ML models 
({\it{e.g.}}, Bayesian networks, decision trees, multiple weak 
classifiers, Markov models, or perhaps scientifically 
derived similarity metrics) will be integrated with 
astronomical taxonomies and ontologies that will enable 
rapid information extraction, knowledge discovery, and 
scientific decision support for real-time astronomical 
research facility operations –- to follow up on the 10-100K 
alertable astronomical events that will be identified each 
night for 10 years by the LSST sky survey. 

The classification broker will include a knowledgebase to 
capture the new labels (tags) that are generated for the 
new astronomical events.  These tags are annotations to 
the events. ``Annotation'' refers to tagging the data and 
metadata content with descriptive terms. For this knowledgebase, 
we envision a collaborative tagging system, called AstroDAS 
(Astronomy Distributed Annotation System) \cite{bose06}. AstroDAS is 
similar to existing science knowledgebases, such as 
BioDAS (biodas.org), 
WikiProteins (www.wikiprofessional.info), 
the Heliophysics Knowledgebase 
(HPKB; www.lmsal.com/helio-informatics/hpkb/), 
and The Entity Describer \cite{good07}.  
AstroDAS is ``distributed'' in the sense that the source 
data and metadata are distributed, and the users are 
distributed.  ``Annotation'' refers to tagging the data and 
metadata content with descriptive terms, which apply to 
individual data granules or to subsets of the data.  
It is a ``system'' with a unified schema for the 
annotation database, where distributed data are perceived 
as a unified data system to the user. One possible 
implementation of AstroDAS could be as a Web 2.0 
scientific data and information 
mashup (=Science2.0).  AstroDAS users will include providers 
(authors) and annotation users (consumers).  Consumers 
(humans or machines) will eventually interact with 
AstroDAS in four ways:

\begin{enumerate}

\item Integrate the annotation database content within their own data portals,
providing scientific content to their own communities of users.

\item Subscribe to receive notifications when new sources are annotated 
or classified. 

\item Use the classification broker as a data integration tool to 
broker classes and annotations between sky surveys, robotic telescopes, 
and data repositories.

\item Query the annotation database (either manually or through web services). 

\end{enumerate}

In the last case, the users include the astronomical event 
message producers, who will want to issue their alerts with 
their best-estimate for the astronomical classification of 
their event.  The classification will be generated through 
the application of ML algorithms to the 
networked data accessible via the VO, in order to arrive 
at a prioritized list of classes, ordered by probability of certainty.
In order to facilitate these science use cases 
(and others not listed here), AstroDAS must have the following features:
(a)~it must enable collaborative, dynamic, distributed
sharing of annotations; 
(b)~it must access databases, data repositories, grids, and web services;
(c)~it must apply ontologies, semantics, dictionaries, annotations, and tags; and
(d)~it must employ data/text mining, ML, and information
extraction algorithms.

\subsection{Collaborative Annotation of Classes}

Machine learning and data mining algorithms, when applied to 
very large data streams, could possibly generate the classification 
labels (tags) autonomously.  Generally, scientists do not 
want to leave this decision-making to machine intelligence 
alone -– they prefer to have human intelligence in the loop also.  
When humans and machines work together to produce the best 
possible classification label(s), this is collaborative 
annotation. Collaborative annotation is a form of 
Human Computation \cite{vonahn04}.  Human Computation refers 
to the application of human intelligence to solve 
complex difficult problems that cannot be solved by 
computers alone.  Humans can see patterns and semantics 
(context, content, and relationships) more quickly, 
accurately, and meaningfully than machines.  
Human Computation therefore applies to the problem of 
annotating, labeling, and classifying voluminous 
data streams.  Of course, the application of autonomous 
machine intelligence (data mining and ML) 
to the annotation, labeling, and classification of 
data granules is also valid and efficacious.  
The combination of both human and machine intelligence 
is critical to the success of AstroDAS as a 
classification broker for enormous data-intensive 
astronomy sky survey projects, such as LSST.

\subsection{A Research Agenda}
\label{s:agenda}

We identify some of the key research activities that 
must be addressed, in order to promote the development 
of a ML-based classification broker for 
petascale mining of large-scale astronomy sky survey 
databases.  Many of these research activities are 
already being pursued by other data mining and 
computational science researchers –- we hope to 
take advantage of all such developments, many of 
which are enabled through advanced next-generation 
data mining and cyber-infrastructure research:

\begin{enumerate}

\item Before the classification labels can be useful, 
we must reach community consensus on the correct set 
of semantic ontological, taxonomical, and 
classification terms.  There are ontologies under 
development in astronomy already –- their completeness, 
utility, and usability need to be researched.

\item Research into user requirements and scientific 
use cases will be required in order that we design, 
develop, and deploy the correct user-oriented 
petascale data mining system.

\item A complete set of classification rules must 
be researched and derived for all possible 
astronomical events and objects.  For objects and 
events that are currently unknown, we need to 
identify robust outlier and novelty detection rules 
and classifiers. These need to be researched and tested.

\item We need to research and collect comprehensive 
sets of training examples for the numerous classes 
that we hope to classify.  With these samples, the 
classification broker will be trained and validated.

\item Algorithms for web services-based (perhaps 
grid-based or peer-to-peer) classification and mining 
of distributed data must be researched, developed, and 
validated.  These mining algorithms should include text 
mining as well as numeric data mining, perhaps an 
integrated text-numeric data mining approach will be 
most effective and thus needs to be researched.

\item User interface and interaction models will need 
to be researched through prototypes and demonstrations 
of the classification broker.

\item Research into the robust integration of the many 
AstroDAS system components will be needed.  
This will require investigation of different modes of 
interaction and integration, such as grids, web services, 
RSS feeds, ontologies (expressed in RDF or OWL), 
linked databases, etc.

\item Deploy a working classification broker on a live 
astronomical event message stream, to research its 
functionality, usefulness, bottlenecks, failure modes, 
security, robustness, and (most importantly) scalability
(from the current few events per night, up to
many tens of thousands of events per night
in the coming decade).  Fortunately, there are such 
event message feeds available today, though on a much 
smaller scale than that anticipated from LSST.

\end{enumerate}

Clearly, this is an ambitious research agenda.  
It will not be fully accomplished in just a year or two.  
It will require several years of research and development.  
This is fortunate, since the most dramatic need for the 
classification broker system for astronomy will come with 
the start-up of LSST sky survey operations in 2016, 
lasting ten years (until 2026).  So, we have a few years 
to get it right, and we will need all of those years to 
complete the challenging research program described above.

\section{Introducing the New Science of Astroinformatics}
\label{s:astroinfo}

As described above, 
today's astronomical research environment is highly focused on the
design, implementation, and archiving of very large sky surveys.
Many projects today ({\em e.g.}, 
Palomar-Quest Synoptic Sky Survey [PQ], 
Sloan Digital Sky Survey [SDSS], and
2-Micron All Sky Survey [2MASS]) plus many more projects in
the near future ({\em e.g.}, LSST, 
Palomar Transient Factory [PTF],
Supernova Acceleration Probe [SNAP], 
Panoramic Survey Telescope And Rapid Response System [Pan-STARRS], and
Dark Energy Survey [DES]) are destined to produce 
enormous catalogs of astronomical sources. The virtual collection of these
gigabyte, terabyte, and (eventually) petabyte catalogs will significantly 
increase science return and enable remarkable new scientific discoveries 
through the integration and cross-correlation
of data across these multiple survey dimensions.
Astronomers will be unable to tap the riches of this data lode without a new 
paradigm for {\it{astroinformatics}} that involves distributed database
queries and data mining across distributed virtual tables of
de-centralized, joined, and integrated sky survey catalogs. The challenges
posed by this problem are daunting, as in most disciplines today that
are producing data floods at prodigious rates.

The development and deployment of the astronomy Virtual Observatory (VO) 
is perceived by some as the solution to this problem. The VO 
provides one-stop shopping
for all end-user data needs, including access to distributed heterogeneous
data, services, and other resources ({\em e.g.}, 
the GRID). Some grid-based data mining
services are already envisioned or in development ({\em e.g.},
GRIST at http://grist.caltech.edu/, 
the Datamining Grid,
and F-MASS at http://www.itsc.uah.edu/f-mass/).
However, processing and mining the 
associated distributed and vast data collections are fundamentally challenging 
since most off-the-shelf data mining systems require the data to be 
downloaded to a single location before further analysis. This imposes serious 
scalability constraints on the data mining system and fundamentally hinders 
the scientific discovery process. If distributed data repositories are to
be really accessible to a larger community, then technology ought to be
developed for supporting distributed data analysis that can reduce, as much as
possible, communication requirements.

The new science of {\it{astroinformatics}}
will emerge from this large and expanding distributed heterogeneous data
environment. We define astroinformatics as {\it{the formalization of 
data-intensive astronomy for research and education}}
\cite{borne2009a, borne2009b}.
Astroinformatics will
borrow heavily from concepts in the fields of bioinformatics
and geoinformatics ({\em i.e.}, GIS = Geographic Information Systems).
The main features of this new science are: it is data-driven, data-centric,
and data-inspired. As bioinformatics represents an entirely new paradigm
for research in the biological sciences, beyond computational biology,
so also does astroinformatics represent a new mode of 
data-intensive scientific research
in astronomy that is cognizant of and dependent on the astronomical flood
of astronomical data that is now upon us. Data mining and knowledge
discovery will become the killer apps for this mode of scientific
research and discovery. Scientific databases will be the ``virtual sky''
that astronomers will study and mine. New scientific understanding will 
flow from the discovered knowledge, which is derived from the
avalanche of information content, which is extracted from the massive data 
collections.

\subsection{Distributed Scientific Data Mining}
\label{s:dsdm}

Distributed data mining (DDM)
of large scientific data collections will become the norm
in astronomy, as the data collections (from the numerous
large sky surveys) become so large
that they cannot all be downloaded to a central
site for mining and analysis. DDM algorithms will be an essential
tool to enable discovery of the hidden knowledge buried among
geographically dispersed heterogeneous databases 
\cite{borne03, borne05, sabine2005, gianella06, dutta07}.

As an example of the potential astronomical research that DDM will 
enable, we consider the large survey databases being produced (now
and in the near future) by various NASA missions. GALEX is producing
all-sky surveys at a variety of depths in the near-UV and far-UV.
The Spitzer Space Telescope is conducting numerous large-area surveys
in the infrared, including regions of sky ({\it{e.g.}}, the Hubble Deep Fields)
that are well studied by the Hubble Space
Telescope (optical), Chandra X-ray Observatory, and
numerous other observatories. The WISE mission (to be launched
circa 2009) will produce an all-sky infrared survey. The 2-Micron
All-Sky Survey (2MASS) has catalogued millions of stars and galaxies
in the near-infrared. Each of these wavebands contributes valuable
astrophysical knowledge to the study of countless classes of objects
in the astrophysical zoo. In many cases, such as the young
star-forming regions within starbursting galaxies, the relevant
astrophysical objects and phenomena have unique characteristics
within each wavelength domain. For example, starbursting galaxies
are often dust-enshrouded, yielding enormous infrared fluxes.
Such galaxies reveal peculiar optical morphologies, occasional
X-ray sources (such as intermediate black holes), and possibly even
some UV bright spots as the short-wavelength radiation leaks through
holes in the obscuring clouds. All of these data, from multiple
missions in multiple wavebands, are essential for a full
characterization, classification, analysis, and interpretation
of these cosmologically significant populations. 

In order to
reap the full potential of scientific data mining, analysis, and discovery
that this distributed data environment enables, it is essential to bring
together data from multiple heterogeneously distributed data sites. 
For the all-sky surveys in particular 
(such as 2MASS, WISE, GALEX, SDSS, LSST),
it is impossible to access, mine, navigate, browse, and analyze these
data in their current distributed state. To illustrate this point,
suppose that an all-sky catalog contains descriptive data for one
billion objects; and suppose that these descriptive data consist
of a few hundred parameters (which is typical for the 2MASS and Sloan
Digital Sky Surveys). Then, assuming simply that each parameter
requires just 2-byte representation, then each survey database will
consume one terabyte of space. If the survey also has a temporal
dimension (such as the LSST, which will re-image each object
1000-2000 times), then massively more data handling is required
in order to mine the enormous potential of the
database contents.  If each of these
catalog entries and attributes requires only one CPU cycle to process
it ({\it{e.g.}}, in a data mining operation), then many teraflops (up to 
petaflops) of computation will be required even for the 
simplest data mining application on  the full contents
of the databases.

It is clearly infeasible,
impractical, and impossible to drag these terabyte (and soon, petabyte) catalogs
back and forth from user to user, from data center to data center,
from analysis package to package, each time someone has a new query
to pose against these various data collections. Therefore,
there is an urgent need for novel DDM algorithms that are
inherently designed to work on distributed data collections.
We are consequently
focusing our research efforts on these problems \cite{gianella06, dutta07}.

\subsection{Beyond the Science}
\label{s:educ}

Before we conclude, it is important to mention how
these scientific data mining concepts are also
relevant to science, mathematics, 
and technical education in our society today \cite{borne2009b}.
The concept of ``Using Data in the Classroom'' is
developing quite an appeal among inquiry-based learning 
proponents\footnote{http://serc.carleton.edu/usingdata/}. Astronomy
data and images in particular 
have a special universal appeal to students, general public,
and all technical experts. 
Student-led data mining projects that access large astronomical databases
may lead to discoveries of new comets, asteroids,
exploding stars, and more. Members of both the LSST and the NVO project 
scientific teams are
especially interested in this type of collaboration among
scientists, data mining experts, educators, and students. The classroom
activities (involving ``cool astronomy data'') are engaging and exciting
to students and thus contribute to the overall
scientific, technical, and mathematical literacy of the nation. 
Astroinformatics enables transparent data sharing, reuse,
and analysis in
inquiry-based science classrooms.
This allows not only scientists, but also students, educators,
and citizen scientists to tackle
knowledge discovery problems in large astronomy
databases for fun and for real.
This integrated research and education 
activity matches well to the objectives of the new CODATA 
ADMIRE 
(Advanced Data Methods and Information technologies 
for Research and Education) initiative
(www.iucr.org/iucr-top/data/docs/codataga2006\_beijing.html).
Students are trained: 
(a)~to access large distributed data repositories;
(b)~to conduct meaningful scientific inquiries into the data;
(c)~to mine and analyze the data;
and (d)~to make data-driven scientific discoveries
\cite{borne2009c}.

\subsection{Informatics for Scientific Knowledge Discovery}
\label{s:summ}

Finally, we close with discussions of BioDAS (the inspiration 
behind AstroDAS) and of the relevance of informatics 
({\it{e.g.}}, Bioinformatics and Astroinformatics) to the 
classification broker described earlier.  
Informatics is {\it{the discipline of organizing, accessing, 
mining, analyzing, and visualizing data for scientific 
discovery}}.  Another definition says informatics is {\it{the 
set of methods and applications for integration of large 
datasets across spatial and temporal scales to support 
decision-making, involving computer modeling of natural 
systems, heterogeneous data structures, and data-model 
integration as a framework for 
decision-making}}\footnote{Downloaded from
http://ag.arizona.edu/srnr/research/wr/breshears/informatics\_UA
on August 23, 2007}. 

Massive scientific data collections impose enormous 
challenges to scientists:  how to find the most relevant 
data, how to reuse those data, how to mine data and 
discover new knowledge in large databases, and how to 
represent the newly discovered knowledge. The bioinformatics 
research community is already solving these problems with 
BioDAS (Biology Distributed Annotation System). 
The DAS provides a distributed system for researchers 
anywhere to annotate (mark-up) their own knowledge 
(tagged information) about specific gene sequences. 
Any other researcher anywhere can find this annotation 
information quickly for any gene sequence. Similarly, 
astronomers can annotate individual astronomical objects 
with their own discoveries. These annotations can be 
applied to observational data/metadata within distributed 
digital data collections. The annotations provide mined 
knowledge, class labels, provenance, and semantic 
(scientifically meaningful) information about the 
experiment, the experimenter, the object being 
studied (astronomical object in our case, or gene 
sequence in the case of the bioinformatics research 
community), the properties of that object, new features 
or functions discovered about that object, its 
classification, its connectiveness to other objects, and so on. 

Bioinformatics (for biologists) and Astroinformatics 
(for astronomers) provide frameworks for the curation, 
discovery, access, interoperability, integration, 
mining, classification, and understanding of digital 
repositories through (human plus machine) semantic 
annotation of data, information, and knowledge. 
We are focusing new research efforts on further development of 
Astroinformatics as: (1)~a new subdiscipline of 
astronomical research (similar to the role of 
bioinformatics and geoinformatics as stand-alone 
subdisciplines in biological and geoscience 
research and education, respectively); and 
(2)~the new paradigm for data-intensive astronomy 
research and education \cite{borne2009a, borne2009b}, 
which focuses on existing 
cyberinfrastructure such as the astronomical Virtual 
Observatory.

\bibliographystyle{plain}

\begin{thebibliography}{}

\bibitem{andreon2000}
Andreon, S., Gargiulo, G., Longo, G., Tagliaferri, R., \& Capuano, N. 2000,
``Wide field imaging - I. Applications of neural networks to object detection 
and star/galaxy classification,''
Monthly Notices of the Royal Astronomical Society, 319, p. 700.

\bibitem{angel90}
Angel, J. R. P., Wizinowich, P., Lloyd-Hart, M., \& Sandler, D. 1990,
``Adaptive Optics for Array Telescopes Using Neural Network Techniques,''
Nature, 348, p. 221.

\bibitem{ball2004}
Ball, N. M., et al. 2004,
``Galaxy Types in the Sloan Digital Sky Survey Using
Supervised Artificial Neural Networks,''
Monthly Notices of the Royal Astronomical Society, 348, p. 1038.

\bibitem{ball2006}
Ball, N. M., Brunner, R. J., Myers, A. D., \& Tcheng, D. 2006,
``Robust Machine Learning Applied to Astronomical Data Sets. 
I. Star-Galaxy Classification of the Sloan Digital Sky Survey
DR3 Using Decision Trees,''
Astrophysical Journal, 650, p. 497.

\bibitem{ball2007a}
Ball, N. M., Brunner, R. J., \& Myers, A. D. 2007,
``Robust Machine Learning Applied to Terascale Astronomical Datasets,''
downloaded from http://arxiv.org/abs/0710.4482.

\bibitem{ball2007b}
Ball, N. M., et al. 2007,
``Robust Machine Learning Applied to Astronomical Data Sets. 
II. Quantifying Photometric Redshifts for Quasars Using
Instance-Based Learning,''
Astrophysical Journal, 663, p. 774.

\bibitem{ballbrunner2009}
Ball, N. M., \& Brunner, R. J. 2009,
``Data Mining and Machine Learning in Astronomy,''
submitted to the International Journal of Physics D,
downloaded from http://arxiv.org/abs/0906.2173.

\bibitem{bazell98}
Bazell, D., \& Peng, Y. 1998,
``A Comparison of Neural Network Algorithms and Preprocesing Methods
for Star-Galaxy Discrimination,''
Astrophysical Journal Supplement, 116, p. 47.

\bibitem{becla06}
Becla, J., et al. 2006, ``Designing a multi-petabyte database 
for LSST,'' downloaded from http://arxiv.org/abs/cs/0604112.

\bibitem{bell05}
Bell, G., Gray, J., \& Szalay, A. 2005, ``Petascale computations 
systems: Balanced cyberinfrastructure in a data-centric world,'' 
downloaded from http://arxiv.org/abs/cs/0701165.

\bibitem{bloom2008}
Bloom, J. S., et al. 2008,
``Towards a Real-time Transient Classification Engine,''
downloaded from http://arxiv.org/abs/0802.2249.

\bibitem{borne01A}
Borne, K. D. 2001, ``Science User Scenarios for a 
VO Design Reference Mission: Science Requirements for Data Mining,'' 
in Virtual Observatories of the Future, San Francisco: Astronomical 
Society of the Pacific, p.333.

\bibitem{borne01B}
Borne, K. D. 2001, ``Data Mining in Astronomical Databases,'' 
in "Mining the Sky."  New York: Springer-Verlag, p.671.

\bibitem{borne03}
Borne, K. D. 2003,
``Distributed Data Mining in the National Virtual Observatory,''
Proceedings of the SPIE, volume 5098, Data Mining and Knowledge Discovery:
Theory, Tools, and Technology V, Bellingham, WA: SPIE, p. 211.

\bibitem{borne05}
Borne, K. D. 2005,
``Data Mining in Distributed Databases for Interacting Galaxies,''
Astronomical Data Analysis Software and Systems XIV, 
A.S.P. Conference Series, San Francisco: Astronomical 
Society of the Pacific, Vol. 347, p. 350.

\bibitem{borne06}
Borne, K. D. 2006, ``Data-Driven Discovery through e-Science 
Technologies,'' in the proceedings of the IEEE conference on 
Space Mission Challenges for Information Technology, 
Washington, DC: IEEE Computer Society.

\bibitem{borne08}
Borne, K. D. 2008, ``A Machine Learning Classification Broker
for the LSST Transient Database,''
Astronomische Nachrichten, 329, p. 255.

\bibitem{borne2008b}
Borne, K. D., Strauss, M. A., \& Tyson, J. A. 2007,
``Data Mining Research with the LSST,''
downloaded from http://adsabs.harvard.edu/abs/2007AAS...21113725B.

\bibitem{borne2009a}
Borne, K. D. 2009,
``Astroinformatics: A 21st Century Approach to Astronomy,''
State of the Profession position paper submitted March 2009 to the
National Academies NRC Astro2010 Decadal Survey Committee
for Astronomy \& Astrophysics,
downloaded from http://arxiv.org/abs/0909.3892.

\bibitem{borne2009b}
Borne, K. D., et al. 2009,
``The Revolution in Astronomy Education: Data Science for the Masses,''
State of the Profession position paper submitted March 2009 to the
National Academies NRC Astro2010 Decadal Survey Committee
for Astronomy \& Astrophysics,
downloaded from http://arxiv.org/abs/0909.3895.

\bibitem{bornechang}
Borne, K. D., \& Chang, A. 2007,
``Data Mining for Extra-Solar Planets,''
Astronomical Data Analysis Software and Systems XVI, 
A.S.P. Conference Series, San Francisco: Astronomical 
Society of the Pacific, Vol. 376, p. 453.

\bibitem{borneast06}
Borne, K. D., \& Eastman, T. 2006, ``Collaborative 
Knowledge-Sharing for E-Science,'' in the proceedings of the 
AAAI conference on Semantic Web for Collaborative Knowledge Acquisition, Menlo Park, CA: AAAI Press, p. 104. 

\bibitem{borne2009c}
Borne, K., Wallin, J., \& Weigel, R. 2009,
``The New Undergraduate Program in Computational and
Data Sciences at GMU,''
Lecture Notes in Computer Science, vol. 5545, pp. 74–83.

\bibitem{bose06}
Bose, R., Mann, R., \& Prina-Ricotti, D. 2006. 
``AstroDAS: Sharing Assertions across Astronomy Catalogues 
through Distributed Annotation,'' in Lecture Notes in 
Computer Science (LNCS), Berlin: Springer, Volume 4145, p. 193.

\bibitem{brunner2001}
Brunner, R. J., Djorgovski, S. G., Prince, T. A., \& Szalay, A. S. 2001,
``Massive Datasets in Astronomy,''
downloaded from http://arxiv.org/abs/astro-ph/0106481.

\bibitem{budavari2007}
Budavari, T., \& Szalay, A. S. 2007,
``Probabilistic Cross-Identification of Astronomical Sources,''
downloaded from http://arxiv.org/abs/0707.1611.

\bibitem{carliles2007}
Carliles, S., Budavari, T., Heinis, S., Priebe, C., \& Szalay, A. 2007,
``Photometric Redshift Estimation on SDSS Data Using Random Forests,''
downloaded from http://arxiv.org/abs/0711.2477.

\bibitem{collister2004}
Collister, A. A., \& Lahav, O. 2004,
``ANNz: Estimating Photometric Redshifts Using Artificial
Neural Networks,''
Publications of the ASP, 116, p. 345.

\bibitem{cortiglioni2001}
Cortiglioni, F., Mahonen, P., Hakala, P., \& Frantti, T. 2001,
``Automated Star-Galaxy Discrimination for Large Surveys,''
Astrophysical Journal, 556, p. 937.

\bibitem{fundplane}
Djorgovski, S., \& Davis, M. 1987,
``Fundamental Properties of Elliptical Galaxies,''
Astrophysical Journal, 313, p. 59.

\bibitem{djorgovski01a}
Djorgovski, S. G., et al. 2000a,
``Searches for Rare and New Types of Objects,''
downloaded from http://arxiv.org/abs/astro-ph/0012453.

\bibitem{djorgovski01b}
Djorgovski, S. G., et al. 2000b,
``Exploration of Large Digital Sky Surveys,''
downloaded from http://arxiv.org/abs/astro-ph/0012489.

\bibitem{djorgovski01c}
Djorgovski, S. G., et al. 2001,
``Exploration of Parameter Spaces in a Virtual Observatory,''
downloaded from http://arxiv.org/abs/astro-ph/0108346.

\bibitem{skicat1994}  
Djorgovski, S. G., Weir, N., \& Fayyad, U. 1994,
``Cataloging of the Northern Sky from the POSS-II 
using a Next-Generation Software Technology,''
Astronomical Data Analysis Software and Systems III, 
A.S.P. Conference Series, San Francisco: Astronomical 
Society of the Pacific, Vol. 61, p. 195.

\bibitem{dutta07}
Dutta, H., Gianella, C., Borne, K., \& Kargupta, H. 2007,
``Distributed Top-K Outlier Detection from Astronomy Catalogs
using the DEMAC System,'' in the proceedings of the
2007 SIAM International Conference on Data Mining, Philadelphia: SIAM, p. 473.

\bibitem{ferreras}
Ferreras, I., et al. 2006,
``A Principal Component Analysis Approach to the Star Formation
History of Elliptical Galaxies in Compact Groups,''
downloaded from http://arxiv.org/abs/astro-ph/0511753.

\bibitem{firth2003}
Firth, A. E., Lahav, O., \& Somerville, R. S. 2003,
``Estimating Photometric Redshifts with Artificial Neural Networks,''
Monthly Notices of the Royal Astronomical Society, 339, p. 1195.

\bibitem{gavri2001}
Gavrishchaka, V. V., \& Ganguli, S. B. 2001,
``Support Vector Machine as an Efficient Tool for High-Dimensional
Data Processing: Application to Substorm Forecasting,''
Journal of Geophysics Research, 106, p. 29911.

\bibitem{gianella06}
Gianella, C., Dutta, H., Borne, K., Wolff, R., \& Kargupta, H. 2006,
``Distributed Data Mining for Astronomy Catalogs,'' in the proceedings of the
2006 SIAM International Conference on Data Mining, Philadelphia: SIAM.

\bibitem{goderya2002}
Goderya, S. N., \& Lolling, S. M. 2002,
``Morphological Classification of Galaxies using Computer Vision and 
Artificial Neural Networks: A Computational Scheme,''
Astrophysics \& Space Science, 279, p. 377.

\bibitem{good07}
Good, B., Kawas, E., \& Wilkinson, M. 2007, 
``Bridging the gap between social tagging and semantic 
annotation: E.D. the Entity Describer,'' downloaded on 
September 26, 2007 from http://precedings.nature.com/documents/945/version/2

\bibitem{gray2002}
Gray, J., et al. 2002,
``Data Mining the SDSS SkyServer Database,''
downloaded from http://arxiv.org/abs/cs/0202014.

\bibitem{howard1975} 
Howard, R. A., North, D. W., \& Pezier, J. P. 1975,
``A new methodology to integrate planetary quarantine requirements 
into mission planning, with application to a Jupiter orbiter,''
Final Report Stanford Research Inst., Menlo Park, CA. 

\bibitem{jeffrey86}
Jeffrey, W., \& Rosner, R. 1986,
``Optimization Algorithms -- Simulated Annealing and Neural Network Processing,''
Astrophysical Journal, 310, p. 473.

\bibitem{kampak08}
Kampakoglou, M., Rotta, R., \& Silk J. 2008,
``Monolithic or Hierarchical Star Formation? A New Statistical Analysis,''
Monthly Notices of the Royal Astronomical Society, 384, p. 1414.

\bibitem{longo2001}
Longo, G., et al. 2001,
``Advanced Data Mining Tools for Exploring Large Astronomical Databases,''
Proceedings of the SPIE, volume 4477, Astronomical Data Analysis, 
Bellingham, WA: SPIE, p. 61.

\bibitem{mahabal2008}
Mahabal, A., et al. 2008, ``Automated Probabilistic Classification
of Transients and Variables,''
Astronomische Nachrichten, 329, p. 288.

\bibitem{mahonen2000}
Mahonen, P., \& Frantti, T. 2000, ``Fuzzy Classifier for Star-Galaxy Separation,''
Astrophysical Journal, 541, p. 261.

\bibitem{mah07}
Mahootian, F., \& Eastman, T. 2008, ``Complementary Frameworks 
of Scientific Inquiry: Hypothetico-Deductive, Hypothetico-Inductive, 
and Observational-Inductive,'' World Futures journal, 65, 61.

\bibitem{sabine2005}
McConnell, S. M., \& Skillicorn, D. B. 2005,
``Distributed Data Mining for Astrophysical Datasets,''
Astronomical Data Analysis Software and Systems XIV, 
A.S.P. Conference Series, San Francisco: Astronomical 
Society of the Pacific, Vol. 347, p. 360.

\bibitem{mcdow04}
McDowell, J. C. 2004, ``Downloading the Sky,'' IEEE Spectrum, 41, p. 35.

\bibitem{mould04}
Mould, J. 2004, ``LSST Followup,'' downloaded from 
http://www.lsst.org/Meetings/CommAccess/abstracts.shtml 
on August 23, 2007.

\bibitem{naim95}
Naim, A., Lahav, O., Sodre, L., \& Storrie-Lombardi, M. C. 1995,
``Automated Morphological Classification of APM Galaxies by
Supervised Artificial Neural Networks,''
Monthly Notices of the Royal Astronomical Society, 275, p. 567.

\bibitem{odewahn92}
Odewahn, S. C., Stockwell, E. B., Pennington, R. L.,
Humphreys, R. M., \& Zumach, W. A. 1992,
``Automated Star/Galaxy Discrimination with Neural Networks,''
Astronomical Journal, 103, p. 318.

\bibitem{odewahn93}
Odewahn, S. C., Humphreys, R. M., Aldering, G., \& Thurmes, P. 1993,
``Star-Galaxy Separation with a Neural Network. 2: Multiple Schmidt
Plate Fields,'' Publications of the ASP, 105, p. 1354.

\bibitem{olmeda2008}
Olmeda, O., Zhang, J., Wechsler, H., Poland, A., \& Borne, K. 2008,
``Automatic Detection and Tracking of Coronal Mass Ejections 
in Coronagraph Time Series,''
Solar Physics, 248, 485.

\bibitem{oyaizu2008}
Oyaizu, H., et al. 2008, 
``A Galaxy Photometric Redshift Catalog for the
Sloan Digital Sky Survey Data Release 6,''
Astrophysical Journal, 674, p. 768.

\bibitem{bohdan2000}
Paczynski, B. 2000, ``Monitoring All Sky for Variability,''
Publications of the ASP, 112, p. 1281. 

\bibitem{philip2002}
Philip, N. S., Wadadekar, Y., Kembhavi, A., \& Joseph, K. B. 2002,
``A Difference Boosting Neural Network for Automated
Star-Galaxy Classification,''
Astronomy \& Astrophysics, 385, p. 1119.

\bibitem{plante04}
Plante, R., et al. 2004, ``VO Resource Registry,'' in the 
proceedings of the ADASS XIII conference, downloaded on 
August 23, 2007 from http://www.us-vo.org/pubs/index.cfm.

\bibitem{qahwaji2008}
Qahwaji, R., Colak, T., Al-Omari, M., \& Ipson, S. 2008,
``Automated Prediction of CMEs Using Machine Learning
of CME Flare Associations,'' Solar Physics, 248, 471.

\bibitem{qin2003}
Qin, D.-M., Guo, P., Hu, Z.-Y., \& Zhao, Y.-H. 2003,
``Automated Separation of Stars and Normal Galaxies Based 
on Statistical Mixture Modeling with RBF Neural Networks,''
Chinese Journal of Astronomy \& Astrophysics, 3, p. 277.

\bibitem{qu2006}
Qu, M., Shih, F. Y., Jing, J., \& Wang, H. 2006,
``Automatic Detection and Classification of Coronal Mass Ejections,''
Solar Physics, 237, p. 419.

\bibitem{rohde05}
Rohde, D. J., et al. 2005,
``Applying Machine Learning to Catalogue Matching in Astrophysics,''
Monthly Notices of the Royal Astronomical Society, 360, p. 69.

\bibitem{rohde06}
Rohde, D. J., Gallagher, M. R., Drinkwater, M. J., \& Pimbblet, K. A. 2006,
``Matching of Catalogues by Probabilistic Pattern Classification,''
Monthly Notices of the Royal Astronomical Society, 369, p. 2.

\bibitem{salzberg95}
Salzberg, S., et al. 1995,
``Decision Trees for Automated Identification of
Cosmic-Ray Hits in the Hubble Space Telescope Images,''
Publications of the ASP, 107, p. 279.

\bibitem{sebok79}
Sebok, W. 1979, ``Optimal classification of images into stars or galaxies 
-- A Bayesian approach,'' Astronomical Journal, 84, p. 1526.

\bibitem{storrie92}
Storrie-Lombardi, M. C., Lahav, O., Sodre, L., \& Storrie-Lombardi, L. J. 1992,
``Morphological Classification of Galaxies by Artificial Neural Networks,''
Monthly Notices of the Royal Astronomical Society, 259, p. 8.

\bibitem{strauss04}
Strauss, M. 2004, ``Towards a Design Reference Mission 
for the LSST,'' downloaded on August 23, 2007 from
http://www.lsst.org/Meetings/CommAccess/abstracts.shtml.

\bibitem{class-x}
Suchkov, A., et al. 2002,
``Automated Object Classification with ClassX,''
downloaded from http://arxiv.org/abs/astro-ph/0210407.

\bibitem{class-x2}
Suchkov, A. A., Hanisch, R. J., \& Margon, B. 2005,
``A Census of Object Types and Redshift Estimates in the SDSS
Photometric Catalog from a Trained Decision-Tree Classifier,''
downloaded from http://arxiv.org/abs/astro-ph/0508501.

\bibitem{szalay2002}
Szalay, A. S., Gray, J., \& VandenBerg, J. 2002,
``Petabyte Scale Data Mining: Dream or Reality?'', in the
Proceedings of the SPIE, volume 4836, Survey and Other
Telescope Technologies and Discoveries, Bellingham, WA: SPIE, p. 333.

\bibitem{tyson04}
Tyson, J. A. 2004, ``The Large Synoptic Survey Telescope: 
Science \& Design,'' downloaded on August 23, 2007 from 
http://www.lsst.org/Meetings/CommAccess/abstracts.shtml. 

\bibitem{vanzella2004}
Vanzella, E., et al. 2004,
``Photometric redshifts with the Multilayer Perceptron Neural Network: 
Application to the HDF-S and SDSS,''
Astronomy \& Astrophysics, 423, p. 761.

\bibitem{vonahn04}
von Ahn, L., \& Dabbish, L. 2004, ``Labeling Images 
with a Computer Game,'' in the proceedings of the SIGCHI 
conference on Human Factors in Computing Systems,
New York: ACM, p.319.

\bibitem{wadadekar2005}
Wadadekar, Y. 2005,
``Estimating Photometric Redshifts Using Support Vector Machines,''
Publications of the ASP, 117, p. 79.

\bibitem{wang2008}
Wang, D., Zhang, Y.-X., Liu, C., \& Zhao, Y.-H. 2008,
``Two Novel Approaches for Photometric Redshift Estimation based on SDSS and 2MASS,''
Chinese Journal of Astronomy \&  Astrophysics, 8, p. 119.

\bibitem{waniak06}
Waniak, W. 2006, ``Removing cosmic-ray hits from CCD images 
in real-time mode by means of an artificial neural network,''
Experimental Astronomy, vol. 21, issue 3, p. 151.

\bibitem{way2006}
Way, M. J., \& Srivastava, A. N. 2006,
``Novel Methods for Predicting Photometric Redshifts 
from Broadband Photometry Using Virtual Sensors,''
Astrophysical Journal, 647, p. 102.

\bibitem{whitmore1984}
Whitmore, B. C. 1984,
``An Objective Classification System for Spiral Galaxies. I.
The Two Dominant Dimensions,''
Astrophysical Journal, 278, p. 61.

\end{thebibliography}

\end{document}